\documentclass[aps,article,twocolumn,longbibliography,citeautoscript,footinbib,superscriptaddress,prb]{revtex4-2}

\pdfoutput=1						
\usepackage{ragged2e}
\usepackage{amsmath,amsfonts,amssymb}	
\usepackage{mathrsfs}					
\usepackage[cal=boondox,scr=boondoxo]{mathalfa}

\usepackage[pdftex]{graphicx}
\usepackage{microtype}
\usepackage{gensymb}
\usepackage{mdframed}
\usepackage[bbgreekl]{mathbbol}
\usepackage{amssymb}
\usepackage{subcaption}
\usepackage[svgnames]{xcolor}
\usepackage[colorlinks=True,linkcolor=DarkRed,citecolor=ForestGreen,urlcolor=MediumBlue, pdfstartview=FitH,bookmarks=False,pdfpagemode=UseNone
]{hyperref}

\usepackage{feynmp-auto}
\usepackage{natbib}
\usepackage{hyperref}
\usepackage[bbgreekl]{mathbbol}

\usepackage{breqn}
\usepackage{caption}
\usepackage{tcolorbox}
\usepackage{listings}

\definecolor{codegreen}{rgb}{0,0.6,0}
\definecolor{codegray}{rgb}{0.5,0.5,0.5}
\definecolor{codepurple}{rgb}{0.58,0,0.82}
\definecolor{backcolour}{rgb}{1,0.8863,0.7961}

\lstdefinestyle{mystyle}{
    backgroundcolor=\color{backcolour},   
    commentstyle=\color{codegreen},
    keywordstyle=\color{red},
    numberstyle=\tiny\color{codegray},
    stringstyle=\color{codepurple},
    basicstyle=\ttfamily\footnotesize,
    breakatwhitespace=false,         
    breaklines=true,                 
    captionpos=b,                    
    keepspaces=true,                 
    numbers=left,                    
    numbersep=5pt,                  
    showspaces=false,                
    showstringspaces=false,
    showtabs=false,                  
    tabsize=2
}

\makeatletter
\let\cat@comma@active\@empty

\newcommand*{\Rint}{%
  \DOTSI
  \mathop{%
    \mathpalette\@LetterOnInt{\mathscr{P}}%
  }%
  \mkern-\thinmuskip 
  \int
}
\newcommand*{\@LetterOnInt}[2]{%
  \sbox0{$#1\int\m@th$}%
  \sbox2{$%
    \ifx#1\displaystyle
      \textstyle
    \else
      \scriptscriptstyle
    \fi
    #2%
  \m@th$}%
  \dimen@=.4\dimexpr\ht0+\dp0\relax
  \ifdim\dimexpr\ht2+\dp2\relax>\dimen@
    \sbox2{\resizebox*{!}{\dimen@}{\unhcopy2}}%
  \fi
  \dimen@=\wd0 %
  \ifdim\wd2>\dimen@
    \dimen@=\wd2 %
  \fi
  \rlap{\hbox to \dimen@{\hfil
    $#1\vcenter{\copy2}\m@th$%
  \hfil}}%
  \ifdim\dimen@>\wd0 %
    \kern.5\dimexpr\dimen@-\wd0\relax
  \fi
}
\makeatother

\newif\ifshowcomments\showcommentstrue

\begin{document}

\title{Anisotropic scattering rates in strain-tuned Sr$_2$RuO$_4$}

\author{Ben Currie}
\affiliation{Department of Physics, King’s College London, Strand, London WC2R 2LS, United Kingdom}
\author{David T. S. Perkins}
\affiliation{Department of Physics, Loughborough University,
Loughborough LE11 3TU, England, United Kingdom}
\author{Evgeny Kozik}
\affiliation{Department of Physics, King’s College London, Strand, London WC2R 2LS, United Kingdom}
\author{ Joseph J. Betouras}
\affiliation{Department of Physics, Loughborough University,
Loughborough LE11 3TU, England, United Kingdom}
\author{J\"org Schmalian}
\affiliation{Institut f\"ur Theorie der Kondensierten Materie, Karlsruher Institut
f\"ur Technologie, 76131 Karlsruhe, Germany}
\affiliation{Institut f\"ur Quantenmaterialien und Technologien, Karlsruher Institut
f\"ur Technologie, 76131 Karlsruhe, Germany}
\date{\today}

\begin{abstract}
Motivated by recent angle-resolved photoemission spectroscopy (ARPES) experiments, we analyze the temperature, frequency, and momentum dependence of the single-particle scattering rate in a model of the $\gamma$-band of Sr$_2$RuO$_4$ under strain, with particular emphasis on the behavior near the Lifshitz transition where the Fermi energy crosses a single Van Hove point. While the scattering rate is only moderately anisotropic at zero strain, we find that it becomes strongly anisotropic at the Lifshitz point. At the lowest energies, we recover the expected universal behavior: the scattering rate varies (ignoring logarithmic corrections) as $\tau^{-1}\sim \omega $ at the Van Hove point and as $\tau^{-1}\sim \omega^{3/2}$ away from it. At higher energies, however, corrections of order $\omega^2$ become important in both regimes. We show that the experimentally observed behavior $\tau^{-1} \sim \omega^{\alpha}$ with $\alpha \approx 1.4(2)$ at the Van Hove point can be quantitatively explained by a superposition of linear and quadratic contributions to the scattering rate, which are comparable in magnitude at the intermediate energies probed by experiment, rather than in terms of a new universal power law. We further predict a distinctive anisotropy, strain dependence, and a non-monotonic frequency dependence of the scattering rate at a Lifshitz transition, all of which may be directly tested in  experiments.
\end{abstract}

\maketitle

\section{Introduction}
\begin{figure}
    \centering
    \includegraphics[width=1.0\linewidth]{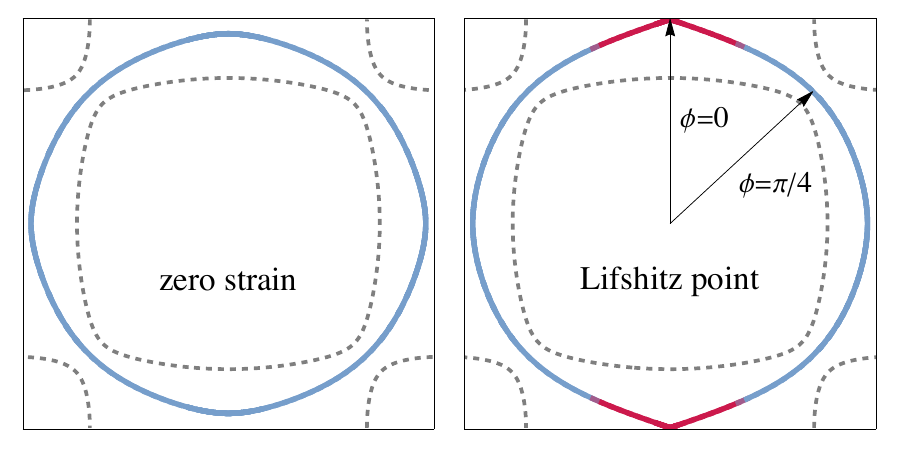}
    \caption{\justifying Fermi surfaces in the first Brillouin zone of Sr$_2$RuO$_4$  at $k_z=0$ for zero strain (left panel) and at the critical strain of the Lifshitz point (right panel). At the Lifshitz point the $\gamma$-sheet of the Fermi surface (shown in color) touches the Van Hove point at in-plane momentum $\boldsymbol{k}=\left(0,\pm\pi\right)$. The part of the Fermi surface near the Van Hove point marked in red refers to the {\em hot} regions, while {\em cold} parts of the Fermi surface, away from the Van Hove point, are shown in blue. The angle $\phi$ marks the azimuthal angle relative to $\left(0,\pi\right)$. Dashed lines indicate the $\alpha$- and $\beta$-sheets of the Fermi surface that are not analyzed in this paper.}
    \label{fig:hot_cold_and_scattering}
\end{figure}

The role of electron-electron interactions in two-dimensional systems
with a Van Hove singularity \cite{VanHove1953} (VHS) in the vicinity of
the Fermi energy has received significant recent interest \cite{Classen_Betouras_2025}. Materials
where the proximity to a Van Hove point is relevant include doped graphene \cite{McChesney2010,Nandkishore2012,Kiesel2012,Wang2012,BlackSchaffer2014},
moiré materials \cite{Shtyk2017,Yuan2019,Isobe2019,Yuan2020,Chandrasekaran2020,Ojajarvi2024},
metallic Kagome superconductors \cite{Hu2022,Kiesel2012B,Park2021,Li2024}, Sr$_{3}$Ru$_{2}$O$_{7}$
in an external magnetic field \cite{Efremov2019}, the bulk of Sr$_{2}$RuO$_{4}$ \cite{Hicks2014,Barber2018,Li2022,Stangier2022}
under uniaxial compressive strain, and the surface of Sr$_{2}$RuO$_{4}$ \cite{Damascelli2000, Chandrasekaran2024}. In particular, the strain-tuned Lifshitz transition of Sr$_{2}$RuO$_{4}$ ~\cite{Lifshitz1960} -- where the Fermi energy crosses a single, quasi two-dimensional VHS -- gives rise to a number of anomalies: At the transition one finds a pronounced maximum of the superconducting transition temperature~\cite{Hicks2014,Steppke2017}; a
giant lattice softening, as observed in a dramatic
reduction of the Young's modulus~\cite{Noad2023}; 
a well established maximum in the electronic entropy, as revealed in measurements of the elasto-caloric effect~\cite{Li2022}; and deviations from the Fermi-liquid $T^{2}$-variation of the resistivity $\rho(T)$~\cite{Barber2018,Yang2023}.
For the lattermost example, transport experiments find an anomalous $T^2\log T$ scaling of $\rho(T)$ at the Lifshitz transition~\cite{Barber2018,Yang2023}, a behavior expected for a Fermi liquid tuned to a Van Hove point\cite{Hlubina1996,Mousatov2020,Stangier2022}. The Fermi surfaces of Sr$_{2}$RuO$_{4}$ at zero strain and at the strain-tuned Lifshitz point are shown in the left and right panels of Fig.~\ref{fig:hot_cold_and_scattering}, respectively, with each consisting of three sheets. Here we focus on the so-called $\gamma$-sheet where the Fermi surface meets the Van Hove point at $\boldsymbol{k}=\left(0,\pm \pi\right)$ (indicated in color in Fig.~\ref{fig:hot_cold_and_scattering}).

The enhanced entropy and compressibility near the Lifshitz transition, as well as the pronounced lattice softening, are directly linked to a broad spectrum of particle-hole, {\em compressive} excitations at small momentum and characterized by the density response: 
\begin{equation}
    {\rm Im}\Pi\left(\boldsymbol{q},\omega\right)=-\frac{m}{2\pi}\left\{ \begin{array}{ccc}
\frac{\omega}{\left|\varepsilon_{{\rm VH}}\left(\boldsymbol{q}\right)\right|} & \,{\rm if}\, & \left|\omega\right|<\left|\varepsilon_{{\rm VH}}\left(\boldsymbol{q}\right)\right|\\
{\rm sign}\left(\omega\right) & \,{\rm if}\, & \left|\omega\right|>\left|\varepsilon_{{\rm VH}}\left(\boldsymbol{q}\right)\right|
\end{array}\right. ,
\label{eq:Pi}
\end{equation}
where $\varepsilon_{{\rm VH}}\left(\boldsymbol{q}\right)=\tfrac{1}{2m}\left(q_{x}^{2}-q_{y}^{2}\right)$
describes the dispersion near the Van Hove point. Notice, this is distinct from the behavior in systems with a spherical Fermi surface, where  ${\rm Im}\Pi\left(\boldsymbol{q},\omega\right)$ is nonzero only for $\left|\omega\right|<v_F\left|\boldsymbol{q}\right|$ with $v_F$ the Fermi velocity. 
The temperature and frequency dependence of the single-particle scattering rate $\tau^{-1}_{\boldsymbol{k}}(\omega,T) $ due to scattering off these soft fermionic compressive excitations is highly anisotropic along the Fermi surface. (For a definition of  $\tau^{-1} $, see Eq.~\eqref{eq:tau}.)
For generic momenta away from the Van Hove point -- called {\em cold} states and marked in blue in Fig.~\ref{fig:hot_cold_and_scattering} --  holds $\tau^{-1}\propto \omega^{3/2}$ or $\propto T^{3/2}$ whichever dominates \cite{Gopalan1992,Stangier2022}. On the other hand, in the immediate vicinity of the Van Hove point -- called {\em hot} states and marked in red in  Fig.~\ref{fig:hot_cold_and_scattering} --  holds $\tau^{-1}\propto \omega /\log(D/\omega)$ or $\propto T /\log(D/T)$ with band width $D$~\cite{Pattnaik1992,Dzyaloshinskii1996,Menashe1999}. These scattering  events provide the dominant contribution to the single-particle self-energy, yet they are not relevant for the electrical resistivity in sufficiently clean systems, since 
scattering processes in a system with a single VHS at the Fermi energy have a small momentum transfer. In the absence of impurities, momentum is relaxed only by umklapp processes, which generally involve large transferred momenta. Hence, only  mild deviation from the conventional Fermi liquid behavior occur in $\rho(T)$. These umklapp processes lead to the above mentioned $T^2\log T$ scaling of the resistivity~\cite{Hlubina1996,Mousatov2020,Stangier2022}.  In contrast, the thermal resistivity is sensitive to small-momentum scattering, and is predicted to obey a $T^{3/2}$ scaling \cite{Stangier2022}.
The  $\propto T/ \log T$,  non Fermi liquid-like behavior of hot carriers has  been evasive and is generically hard to determine. 

Recent angular-resolved photoemission spectroscopy (ARPES) measurements~\cite{Hunter2025} found evidence that the quasi-particle spectral properties, specifically the single-particle scattering rate $\tau^{-1}$, are strongly strain dependent.  At generic values of strain, the frequency dependence of $\tau^{-1}$  was found to obey the canonical Fermi liquid scaling $\tau^{-1}(\omega) \sim \omega^\alpha$ with $\alpha \approx 2$, similar to well established quasiparticles seen at zero strain~\cite{Mackenzie1996,Damascelli2000,Ingle2005,Iwasawa2010,Tamai2019}. On the other hand, close to the critical strain of the Lifshitz transition an anomalous scaling $\alpha=1.4(2)$ was deduced right at the Van Hove  point in the Brillouin zone.  The ARPES experiments were conducted at a temperature $T=11 ~\mathrm{K}$, a regime below the coherence temperature $T_{\rm FL}\sim 25\cdots 40 \,{\rm K}$ of the unstrained samples~\cite{Mackenzie2003,Hussey1998,Katsufuji1996}. 

While theory predicts a singular scattering, the exponent $\alpha=1.4(2)$ obtained in Ref.~\cite{Hunter2025} is distinct from the expected value of $\alpha = 1$. It raises the question of whether the scattering is indeed due to the broad continuum of compressive modes of Eq.~\eqref{eq:Pi}, or whether one has to invoke additional interactions to explain this behavior such as collective spin fluctuations, observed in inelastic neutron scattering experiments~\cite{Sidis1999,Braden2002,Servant2002,Braden2004,Iida2011,Jenni2021}. 
In this context it is important to clarify whether the temperature and energy regimes probed in the ARPES measurements of Ref.~\cite{Hunter2025} can indeed be considered as being in the ultimate low-$T$ regime.  This requires a quantitative analysis of the single-particle scattering processes that is based on a realistic modeling of the electronic structure. 
Clarifying which scattering processes dominate on the Fermi surface is also a key step toward identifying the mechanism responsible for superconductivity.

In this paper, we present a detailed analysis of the single-particle scattering rate in a model of the $\gamma$-band of Sr$_2$RuO$_4$. The analysis is based on a model for the strain dependence of the electronic structure  that has previously been used to successfully describe both the elastocaloric effect~\cite{Li2022} and the softening of the Young's modulus~\cite{Noad2023}. We then perform a perturbative calculation of the scattering rate up to second-order in a local effective Hubbard interaction $U$, recovering the expected universal low-temperature behavior discussed above. Our results further demonstrate that the temperatures and energies at which the ARPES measurements of Ref.~\cite{Hunter2025} were performed do not yet correspond to this universal low-energy, low-temperature regime. In particular, the apparent exponent $\alpha \approx 1.4(2)$ can be understood within our framework as arising from a combination of a quadratic Fermi-liquid contribution
and the hitherto elusive linear behavior associated with the VHS,
\begin{equation}
\tau^{-1}= A \omega + B\omega^2.
\end{equation}
As it is hard to determine logarithmic effects in the numerical analysis, we do not distinguish between $\omega$,  $\omega \log \omega$, or $\omega / \log \omega$. The same is true for the $T$-dependence of the scattering rate. These logarithmic effects are, however, discussed in Appendix~\ref{app:AppendixA} in more detail, where we also comment on the fact that the quantitative importance of the logarithmic corrections are rather small.

Likewise, we find that, in the temperature and frequency range relevant to Ref.~\cite{Hunter2025}, the $\omega^{3/2}$ dependence of the scattering rate for states away from the Van Hove point is significantly affected by $\omega^2$ corrections. This follows from the comparatively small phase space available for processes giving rise to a $\omega^{3/2}$ rate. Our analysis thus explains  the puzzling observations of Ref.~\cite{Hunter2025} and may also aid the interpretation of other probes of the anisotropic scattering rate, such as electronic Raman scattering. In addition, we expect a $T^{3/2}$ dependence of the scattering rate in thermal transport, and hence a thermal conductivity scaling as $\kappa(T)\sim T^{-1/2}$, only for $T< 10\,{\rm K}$.

 \section{The Model}
We focus our analysis on the role of the $\gamma$-band of Sr$_2$RuO$_4$ (Fermi surface sheet in Fig.~\ref{fig:hot_cold_and_scattering} marked in color) that is dominated by Ru-$4d_{xy}$ states. We note there are two additional Fermi surface sheets (marked by dashed lines in Fig.~\ref{fig:hot_cold_and_scattering}), referred to as the $\alpha$- and $\beta$-sheets, that are predominantly made up of coupled Ru-$4d_{xz}$ and $4d_{yz}$ orbitals. The contribution of the $\alpha$- and $\beta$-sheets to the overall density of states is comparatively small~\cite{Mackenzie2003}. More importantly, it is the $\gamma$-sheet where the Fermi energy crosses the VHS upon applying uniaxial stress. A discussion of the role of the $\alpha$- and $\beta$-sheets for the kinematics of scattering processes in the electrical resistivity can be found in Ref.~\cite{Stangier2022}. For the remaining $\gamma$-band
we use the  Hubbard model
\begin{equation}
H=\sum_{\boldsymbol{k},\sigma} \varepsilon_{\boldsymbol{k}}^{\null} c_{\boldsymbol{k}\sigma}^{\dagger}c_{\boldsymbol{k}\sigma}^{\null} + U \sum_{i} n_{i\uparrow}^{\null} n_{i\downarrow}^{\null},
\end{equation}
 where the  strain-dependent electronic dispersion  has the form~\cite{Li2022,Noad2023}:
\begin{eqnarray}
    \varepsilon(\boldsymbol{k}) &=& -2t_x\cos(k_x)-2t_y\cos(k_y)\nonumber \\
    &-&4t'\cos(k_x)\cos(k_y)-\mu
    \label{eq:tight_binding}
\end{eqnarray}
with hopping parameters $t_x = t_0(1-\alpha \epsilon)$, $t_y = t_0(1+\alpha \nu_{xy}\epsilon)$, $t' = t_0'(1-\alpha/2(1-\nu_{xy})\epsilon)$ and $\mu = 1.48t_0$, and where the numerical values are taken as
$\alpha=7.604$, $t_0 = 0.119 ~\mathrm{eV}$, $t_0' = 0.392t_0$ and with Poisson ratio $\nu_{xy} \approx 0.51 $, obtained from  the elastic constants from Ref.~\cite{Ghosh2021}. The uniaxial strain parameter $\epsilon=\partial_x u_x$ with displacement vector $\boldsymbol{u}$, controls the degree of anisotropy of the Fermi surface, and tunes a single Van Hove point to the Fermi surface at the critical strain value $\epsilon_{\mathrm{VH}} \approx -0.44\%$. This is very close to the value at which the Lifshitz transition was observed experimentally in $\mathrm{Sr}_2\mathrm{Ru}\mathrm{O}_4$ \cite{Barber2019,Barber2019_correlations,Sunko2019}. The tight-binding  parametrization  of Eq.~\eqref{eq:tight_binding} follows from fits to ARPES results of Ref.~\cite{Burganov2016} at zero strain, $\epsilon=0$, and led to excellent quantitative agreement for strain dependence of the elastocaloric effect and the Young's modulus~\cite{Li2022,Noad2023}.  Effects due to the dispersion along the $c$-direction are small, in particular near the Van Hove point where three-dimensional effects are only relevant below $2-4\,{\rm K}$~\cite{Mackenzie2003,Bergemann2003}.

The  retarded single-particle self-energy at second-order in $U$ and at finite temperature $T$ is expressed as: 
\begin{widetext}
\begin{equation}\label{self energy}
    \Sigma_{\boldsymbol{k}}(\omega,T) =-U^2\int \frac{d^2\boldsymbol{q}}{(2\pi)^2}\frac{d^2 \boldsymbol{k}_1}{(2\pi)^2}\frac{[n_F(\varepsilon_{\boldsymbol{k}_1})-n_F(\varepsilon_{\boldsymbol{k}_1+\boldsymbol{q}})][n_B\left(\varepsilon_{\boldsymbol{k}_1+\boldsymbol{q}}-\varepsilon_{\boldsymbol{k}_1}\right)+n_F(-\varepsilon_{\boldsymbol{k}-\boldsymbol{q}})]}{\varepsilon_{\boldsymbol{k}_1+\boldsymbol{q}}-\varepsilon_{\boldsymbol{k}_1} +\varepsilon_{\boldsymbol{k}-\boldsymbol{q}}-\omega-i0^+ },
\end{equation}
\end{widetext}
where $n_F$ and $n_B$ are the Fermi and Bose functions, respectively and the integration is over the first Brillouin zone. From this, the quasi-particle scattering rate $\tau^{-1}_{\boldsymbol{k}}(\omega,T)$ is extracted via  
\begin{equation}
    \tau^{-1}_{\boldsymbol{k}}(\omega,T) = -2 Z_{\boldsymbol{k}}^{\null}(T) \mathrm{Im}\Sigma_{\boldsymbol{k}}^{\null}(\omega,T),
    \label{eq:tau}
\end{equation}
where the quasi-particle weight is defined as
\begin{equation}
    Z_{\boldsymbol{k}}(T)  = \left(1-\frac{\partial \mathrm{Re}\Sigma_{\boldsymbol{k}}(\omega,T)}{\partial \omega}\bigg|_{\omega=0}\right)^{-1}.
    \label{eq:qpresidue}
\end{equation}
We will also use the notation 
\begin{equation}
  \Gamma_{\boldsymbol{k}}(\omega,T) = -\mathrm{Im}\Sigma_{\boldsymbol{k}}(\omega,T).  
  \label{eq:Gamma}
\end{equation}
Note that the frequency dependence of the scattering rate is determined entirely by $\Gamma_{\boldsymbol{k}}(\omega,T)$, while the temperature and momentum dependence are also determined by $Z_{\boldsymbol{k}}(T)$.
In what follows we will primarily focus on $\Gamma_{\boldsymbol{k}}(\omega,T)/U^2$ as it does not rely on knowledge of the  value of the effective interaction $U$. In Eq.~\eqref{eq:scatt_rate_exp} below we will also comment on the quasiparticle weight and the scattering rate $\tau^{-1}_{\boldsymbol{k}}(\omega,T) $.

Before presenting the numerical results for the self-energy, we comment on the appropriateness of employing second-order perturbation theory. Sr$_2$RuO$_4$ is a correlated material with  complex interactions that couple spin, charge, and orbital excitations at high energies~\cite{Zingl2019,Suzuki2023}. Numerical renormalization group calculations~\cite{Kugler2020} show that, at low energies $T < T_{\rm FL}$, these orbital excitations are strongly screened, explaining the observed emergence of a good Fermi liquid, see Ref.~\cite{Mackenzie2003}. Our analysis is therefore meaningful in this low-energy, low-temperature regime, where such screening has taken place and the phase-space arguments of Fermi-liquid theory apply. In this regime, second-order perturbation theory is expected to provide the dominant contribution to the scattering rate, provided the system does not develop an instability towards an ordered state. Higher-order processes are then expected to renormalize the overall coefficient in the scattering rate without significantly altering its frequency or temperature dependence. An analysis based on Random Phase Approximation is expected to provide exactly the same qualitative results. In this sense, the interaction parameter $U$ of our analysis should be viewed as an effective interaction, analogous to a Landau parameter. Below, we briefly comment on the appropriate choice of $U$. Of course, at the lowest energies, some instability is expected, e.g. towards a superconducting state. For a discussion of the superconducting instability in systems with a single Van Hove point, see Ref.~\cite{Ojajarvi2024}. We will consider the regime above the onset of superconductivity but still low in energy.

\section{Numerical results}

\subsection{Anisotropy of the scattering rate}
\begin{figure}[t]
    \centering
    \includegraphics[width=1.0\linewidth]{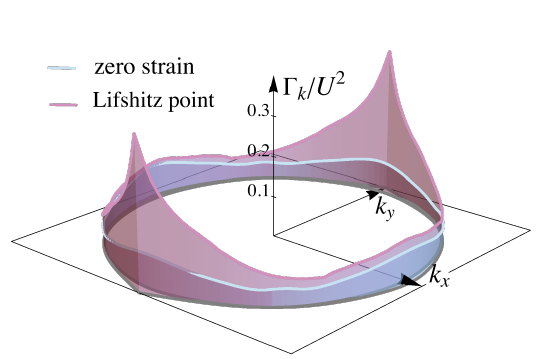}
    \caption{\justifying Anisotropy of the zero-frequency scattering rate $\Gamma_{\boldsymbol{k}_F}/U^2$ on the Fermi surface at $T = 9.9$ K for zero strain (light blue) and the critical strain of the Lifshitz transition (pink). The base area marks the first Brillouin zone boundary and $\Gamma_{\boldsymbol{k}_F}/U^2$ is measured in units of $10^{-2} \text{ eV}^{-1}$. While the scattering rate at zero strain is moderately anisotropic, a pronounced anisotropy with a sharp peak at the Van Hove point emerges at the Lifshitz transition.}
    \label{fig:3dplot_scatt}
\end{figure}

Before we discuss the frequency and temperature dependencies of the scattering rate, we briefly show the anisotropy of $\Gamma_{\boldsymbol{k}}(\omega=0,T)$ at low $T$ along the Fermi surface (Fermi line).
Fig.~\ref{fig:3dplot_scatt} illustrates the momentum dependence of the zero-frequency scattering rate along the Fermi surface, both at zero strain and at the Lifshitz transition. At zero strain (light-blue curve),  $\Gamma_{\boldsymbol{k}}$ exhibits only a moderate angular variation: although the Fermi surface is already anisotropic, no special points dominate the phase space for small-momentum particle-hole excitations. Consequently, the scattering rate remains of comparable magnitude around the entire Fermi contour.

In contrast, at the Lifshitz transition (pink curve), the anisotropy becomes dramatically enhanced. A sharp and narrow peak develops at the Van Hove momenta ($\phi=0$ and symmetry-related points), reflecting the large phase space for scattering off the broad continuum of compressive particle-hole excitations described by Eq.~\eqref{eq:Pi}.  As a result, quasiparticles at the Van Hove point -- our “hot” states -- experience a scattering rate that is significantly larger than elsewhere on the Fermi surface. Away from the VHS (e.g., near $\phi=\pi/4$ or $\phi=\pi/2$ etc.), the scattering rate remains much smaller. These “cold” regions feel only the reduced phase space associated with conventional processes involving finite curvature of the Fermi surface. The result is the pronounced hot–cold dichotomy familiar from systems tuned to a single two-dimensional VHS.

The strong enhancement of $\Gamma_{\boldsymbol{k}}$ near the Van Hove point is closely related to the distinct frequency and temperature dependence discussed later in this section: linear-in-$T$ scattering of hot carriers and $T^{3/2}$ behavior in cold regions. Moreover, the sharp anisotropy shown in Fig.~\ref{fig:3dplot_scatt} provides a direct spectroscopic signature of approaching the Lifshitz transition and is consistent with the strong strain dependence inferred from the ARPES measurements of Ref.~\cite{Hunter2025}.

\subsection{Temperature dependence of $\Gamma$}
\begin{figure}
    \centering
    \includegraphics[width=0.95\linewidth]{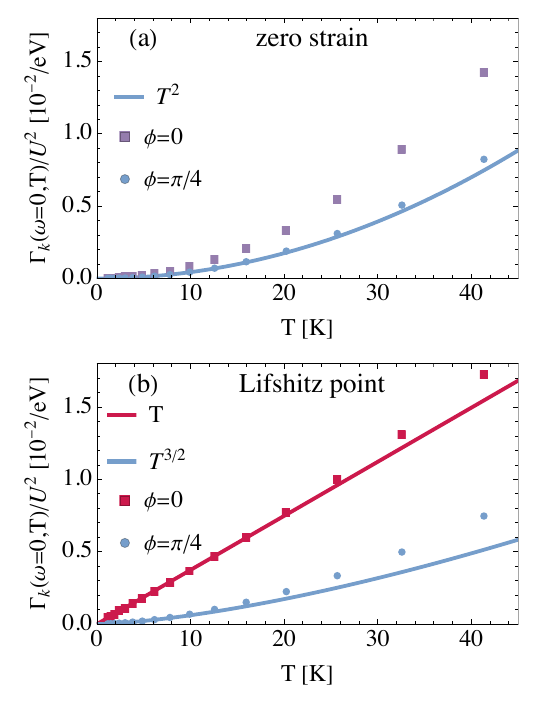} 
    \caption{\justifying Temperature dependence of the imaginary part of the zero-frequency self-energy, 
    $\Gamma_{\boldsymbol{k}}(0,T)$, for cold and hot momenta on the Fermi surface: $\phi = 0$ corresponds to the Van Hove point while $\phi = \pi/4$ is away from it. \textbf{(a):} Zero strain displaying low-$T$ quadratic Fermi liquid behavior with all states being cold. \textbf{(b):} Critical strain exhibiting $T$-linear behavior at $\phi = 0$ (hot) and $T^{3/2}$ behavior away from the Van Hove point with $\phi = \pi/4$ (cold).}
    \label{fig:T_dep_gamma_combined}
\end{figure}

The temperature dependence of the imaginary part of the self-energy at zero frequency, $\Gamma_{\boldsymbol{k}}(\omega=0,T)$, is shown both at zero strain and at the strain value of the Lifshitz transition in the top and bottom panel of Fig.~\ref{fig:T_dep_gamma_combined}, respectively. At zero strain, the Fermi surface has no special features, leading to the canonical Fermi liquid dependence $\Gamma \sim T^2$ at the lowest temperatures at all points on the Fermi surface. For the single-particle self-energy one further expects additional logarithmic corrections to the $T^2$ behavior due to co-linear scattering~\cite{Hodges1971,Bloom1975,Coffey1993}. As this regime is not of our primary concern, we have not attempted to analyze the quantitative magnitude of these corrections. In distinction, at the critical strain, the self-energy acquires a substantial momentum dependence, and exhibits anomalous $\Gamma \sim T$  scalings with temperature at the Van Hove point and $\Gamma\sim T^{3/2}$ away from it. While a slightly better fit to the data can be achieved with an additional term $\sim T \log(D/T)$ included, the distinction is hardly visible and won't be included in our subsequent discussion.
As discussed earlier, these exponents arise from scattering involving the particle-hole excitations of Eq.~\eqref{eq:Pi}~\cite{Pattnaik1992,Gopalan1992,Dzyaloshinskii1996,Menashe1999,Stangier2022}. 
For cold parts of the Fermi surface (marked in blue in Fig.~\ref{fig:hot_cold_and_scattering}) the dominant processes giving rise to these power-laws  are scattering events where a pair of cold and a hot momentum states scatter into another pair cold and another hot momentum states. We refer to these processes as 
ch $\to$ ch.  On the other hand, for the hot parts of the Fermi surface located near the Van Hove point (marked in red in Fig.~\ref{fig:hot_cold_and_scattering}), the scattering rate is caused by pair of hot momentum states scattering into a second pair of hot states. We refer to these events as hh $\to$ hh. Below we will analyze these distinct contributions in detail.

\subsection{Frequency dependence of $\Gamma$}
\begin{figure}
    \centering
    \includegraphics[width=1.0 \linewidth]{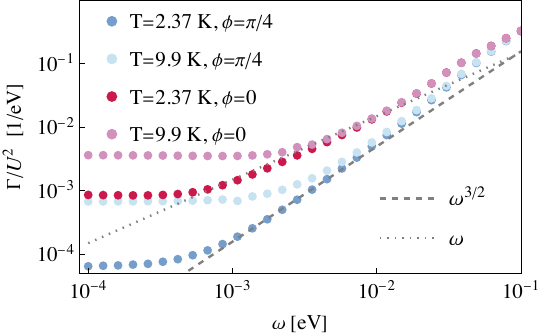}
    \caption{\justifying Log-log plot of the frequency dependence of the imaginary part of the self-energy
    at the critical strain.}
    \label{fig:critical_strain_frequency_dependence_of_gamma}
\end{figure}

\begin{figure}
    \centering
    \includegraphics[width=0.9\linewidth]{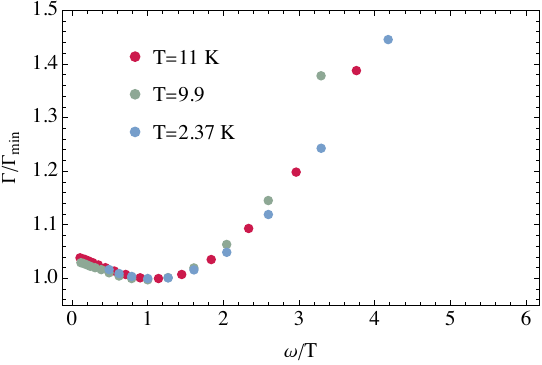}
    \caption{\justifying Scattering rate $\Gamma$ at the Lifshitz point and for the momentum at the Van Hove point as function of frequency for different temperatures. To facilitate the comparison of different $T$, $\Gamma$ has been scaled by the value at the  local minimum and is plotted as function of $\omega/T$. The physical origin of the non-monotonic dependence of the scattering rate is further discussed in Appendix~\ref{app:AppendixA} }
    \label{fig:omega_dep_gamma_linscale}
\end{figure}

For a more direct comparison with the experiment of Ref.\cite{Hunter2025}, which reports the scattering-rate at constant temperature $T=11~\mathrm{K}$ at the Lifshitz transition, we now turn to the frequency dependence of $\Gamma$. The imaginary-part of the self-energy as a function of frequency is shown in Fig.~\ref{fig:critical_strain_frequency_dependence_of_gamma} for two different temperatures $T=2.37 ~\mathrm{K}$ and $T= 9.9~ \mathrm{K}$. At $T=2.37 ~\mathrm{K}$ and within the low-energy window $10^{-3}~\mathrm{eV}$ to $10^{-2}~\mathrm{eV}$, we observe the same scaling exponents seen in the temperature dependence in Fig.~\ref{fig:T_dep_gamma_combined}. Meanwhile, at even lower energies $\omega < 10^{-3} \mathrm{eV}$, we enter the frequency independent regime, where $\Gamma$ depends weakly on frequency but strongly on temperature. On the other hand, at the higher temperature, the expected exponents are not present in any appreciable energy-window due to the onset of the frequency-independent regime at a  higher energy scale, before the universal low-energy regime is reached. Instead, at temperature $T=9.9~ \mathrm{K}$ and at the Van Hove point ($\phi=0$), we find an apparent power-law scaling with a larger exponent of approximately $3/2$ between energies of order $10^{-2} ~\mathrm{eV}$ and $10^{-1}~ \mathrm{eV}$. This is the same power law behavior seen in the ARPES experiment, which obtained an exponent $1.4(2)$ by fitting their ARPES data in the frequency range [-50,10] meV, which is roughly the same range in which we observe this apparent exponent. In Fig.~\ref{fig:omega_dep_gamma_linscale} we also show the frequency dependence at the Van Hove point on a linear scale. Remarkably, at finite temperature, the scattering rate $\Gamma_{\boldsymbol{k}}(\omega,T)$ at the Van Hove point displays a non-monotonic frequency dependence, highlighting the unusual single-particle dynamics at a Lifshitz point. In Appendix~\ref{app:AppendixA} we demonstrate that the scattering rate attains a minimum at $\omega \sim T$, arising from the competing effects of thermal and quantum excitations of the compressive mode described by Eq. (1). At $T=0$, the scattering rate rises, as expected, monotonic with $\omega$.

\subsection{Origin of the apparent high-frequency exponent}
To understand the origin of the putative high frequency power law, we can analyze  the role of the different scattering processes involving hot states near and cold states away from the Van Hove point individually. Since we are considering a single connected Fermi surface, the boundary between the hot and cold regions is somewhat arbitrary: our choice is shown in Fig.~\ref{fig:hot_cold_and_scattering}. Hot parts corresponds to the segment of the Fermi line near $(0,\pm \pi)$ that are essentially straight lines.
Concentrating on the behavior at the Van Hove point, there are several contributions to the scattering rate: (i) hh $\to$ hh  processes that are dominant at low $T$ and yield $\Gamma \sim T$. (ii) ch $\to$ ch processes which are associated with a small transferred momentum yet, come with a large phase space. These events yield $\Gamma \sim T^2$. (iii) hh $\to$ ch  processes that have a rather small phase space since, with the exception of the states close to the boundary between the hot and cold regions,  these events require the transferred momentum to be  small (h $\to$ h) and large (h $\to$ c) at the same time.   (iv) ch $\to$ cc  events that also come with a small phase space, yet play a role in the electrical resistivity as the momentum transfer is large. Notice, since we are considering the self energy at the Van Hove point, we always have at least one hot state.

The decomposition of the second-order self-energy into these different contributions is achieved as follows. We rewrite the Green's function as 
\begin{equation}
    G_{\boldsymbol{k}} = G_{{\rm h},\boldsymbol{k}} + G_{{\rm c},\boldsymbol{k}}= \theta_{\rm h}(\boldsymbol{k})G_{\boldsymbol{k}} + \theta_{\rm c}(\boldsymbol{k})G_{\boldsymbol{k}}
\end{equation}
where the support functions $\theta_{\rm h}(\boldsymbol{k})$ ($\theta_{\rm c}(\boldsymbol{k})$) are unity for momenta $\boldsymbol{k}$ in the hot (cold) regions of the Brillouin zone, and zero otherwise.
The three internal Green’s functions in the self-energy diagram can each be either hot or cold: these possibilities comprise the four channels listed above. For instance, all three Green's function  in the hh $\to$ hh contribution are hot, while the ch $\to$ ch is the sum of the three contributions with a single internal $G_h$.

 \begin{figure}
    \centering
    \includegraphics[width=1.0\linewidth]{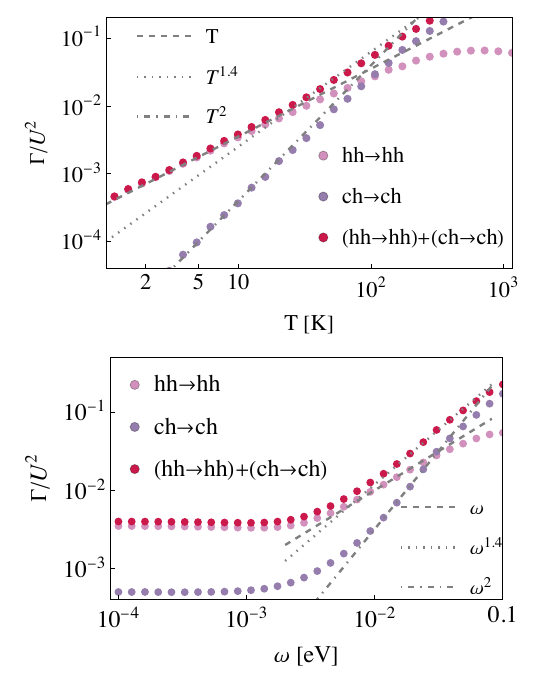}
    \caption{\justifying Contributions to the scattering rate as function of temperature for $\omega=0$ (upper panel) and frequency for $T=11\,{\rm K}$ (lower panel) at the Van Hove momentum and at the Lifshitz transition. We show contributions due to hh $\to$ hh and ch $\to$ ch scattering processes; their sum (shown in red) is indistinguishable from the total scattering rate (not shown). At very low frequencies,  hh $\to$ hh processes dominate and yield a linear behavior, in the intermediate regime ch $\to$ ch processes become comparable and behave like $T^2$ or $\omega^2$. No pure power-law behavior occurs. The sum of different contributions may only appear as a  power-law with an intermediate exponent. }
    
    \label{fig:Two_FSs_Gamma}
\end{figure}

The self-energies arising from only hh $\to$ hh as well as ch $\to$ ch processes are shown in Fig.~\ref{fig:Two_FSs_Gamma}. We see that hh $\to$ hh events are dominant at low temperatures and give a linear-in-temperature scaling, while the contribution from ch $\to$ ch scattering scales with  temperature quadratically and thus becomes sizable  at higher $T$. Remarkably, the sum of these two contributions agrees perfectly with the full self-energy  (not shown in Fig.~\ref{fig:Two_FSs_Gamma}) that includes all possible scattering processes. Hence, these two processes are dominant, and  all others can be neglected. This analysis shows in detail  how the putative intermediate exponent $\alpha=1.4(2)$  seen in Ref.~\cite{Hunter2025} results from the crossover between the $T$ and $T^2$ scalings. We also note that the non-monotonic frequency dependence of the scattering rate, discussed in the previous subsection and in Appendix~\ref{app:AppendixA}, only occurs for hh $\to$ hh processes and is absent in the ch $\to$ ch channel. 

The experiment of Ref.~\cite{Hunter2025} finds $\tau^{-1}_{\mathrm{exp}} = 4 \pm 1~\mathrm{meV}$ for the scattering rate of Eq.~\eqref{eq:tau}  at the Lifshitz point, for  $T = 11~\mathrm{K}$, and at a frequency $\omega = -5~\mathrm{meV}$. 
For the same parameters 
we find $-\mathrm{Im}\Sigma (-5~\mathrm{meV}) = 8.0(2)\tilde{U}^2~\mathrm{meV}$ and $\partial_\omega \mathrm{Re}\Sigma |_{\omega = 0} = -6.1(2) \tilde{U}^2$, where $U = \tilde{U}~\mathrm{eV}$ defines the dimensionless interaction strength $\tilde{U}$. Hence within second-order perturbation theory we could choose a value of $U$ such that 
\begin{equation}
    \tau^{-1}_{\mathrm{exp}} = \frac{16 \tilde{U}^2}{1+6.1\tilde{U^2}} ~\mathrm{meV}.
    \label{eq:scatt_rate_exp}
\end{equation}
For $\tilde{U}\sim 1$ this gives the correct order of magnitude  $\tau^{-1}_{\mathrm{exp}}\sim 2.25~\mathrm{meV}$, yet $\tilde{U}\sim 1$ with a bandwidth or order $1\,{\rm eV}$ is clearly beyond the regime where we can reliably use second-order perturbation theory. 
As discussed earlier, while we expect the power-laws in frequency and temperature to be appropriately reproduced, the precise magnitude of the scattering rate is beyond the accuracy of the second-order perturbation theory. This can already be seen from the fact that $\tau^{-1}_{\mathrm{exp}} = 4~\mathrm{meV}$ cannot be reproduced for any value of $\tilde{U}$, clearly an indication that higher-order processes will change the $\tilde{U}$-dependence in Eq.~\eqref{eq:scatt_rate_exp}. 
Precisely the fact that this estimate yields the correct order of magnitude does seem to suggest that our approach reasonably describes the dominant scattering events of the material in the low-$T$ regime.


\section{Conclusions}
We presented a detailed theoretical analysis of the quasiparticle scattering rate in the $\gamma$-band of strained Sr$_2$RuO$_4$ in the vicinity of the uniaxial-strain-induced Lifshitz transition. Based on a realistic tight-binding model and second-order perturbation theory in an effective local interaction, we find at the Lifshitz point a strong anisotropy of the scattering rate in momentum space and  confirm the universal low-energy behavior associated with a single VHS~\cite{Pattnaik1992,Gopalan1992,Stangier2022}: at the hot spots the scattering rate varies linearly in frequency or temperature, while away from these points we recover the expected $\omega^{3/2}$ or $T^{3/2}$ scaling.

Our results provide a natural explanation for the apparent non-Fermi-liquid exponent $\alpha \approx 1.4(2)$ observed in recent ARPES experiments~\cite{Hunter2025}. We find strong evidence that this intermediate-energy behavior does not reflect new universal physics, but instead arises from a crossover regime in which the linear-in-$\omega$ scattering of hot quasiparticles near the Van Hove point and the quadratic Fermi-liquid–like scattering, involving both hot and cold regions of the Fermi surface, are of comparable magnitude. This competition leads to the effective intermediate exponent reported in Ref.~\cite{Hunter2025}. Nevertheless, the overall energy dependence observed experimentally strongly suggests that, sufficiently close to the Lifshitz critical point, the true low-energy behavior is governed by a linear-in-energy scattering rate.

For states away from the Van Hove point, we find that the comparatively narrow phase space for processes giving rise to the $\omega^{3/2}$ behavior restricts its observability to temperatures below $\approx 10\,{\rm K}$ or energies below 10 meV. In particular, we expect the emergence of a thermal conductivity scaling $\kappa \sim T^{-1/2}$, discussed in Ref.~\cite{Stangier2022}, only for $T \lesssim 10\,{\rm K}$. These results indicate that, aside from the resistivity measurements of Ref.~\cite{Barber2018}, current experiments have not yet accessed the true universal low-energy regime where collision rates are controlled by the VHS and the associated Lifshitz transition. Our work therefore provides quantitative predictions for the strong anisotropy and strain dependence of the scattering rate that can be tested in future spectroscopic and thermal-transport studies.

A further outcome of our analysis is the identification of scattering off the broad continuum of compressive particle-hole excitations described by Eq.~\eqref{eq:Pi} as the dominant low-energy scattering mechanism near the Lifshitz transition. These same excitations are responsible for both the pronounced lattice softening~\cite{Noad2023} and the enhanced normal-state entropy~\cite{Li2022} observed experimentally at the transition, highlighting a common origin of the associated anomalies.

With regard to the superconducting state, the entropy maximum as a function of strain in the normal state is replaced by a minimum below $T_c$~\cite{Li2022}, which has been interpreted as strong evidence that electronic states near the Van Hove point become gapped below $T_c$~\cite{Palle2023}. At the very least, this implies that the compressive particle-hole excitations described by Eq.~\eqref{eq:Pi} must be included in any analysis of the pairing mechanism in Sr$_2$RuO$_4$.

\acknowledgements We are grateful to F. Baumberger, E. Berg, F. Henssler, A. Hunter, M. Le Tacon,  A. P. Mackenzie, P. H. McGuinness, G. Palle, B. J. Ramshaw, and A. Tamai  for helpful discussions. This work was supported by the UK Engineering and Physical Sciences Research Council through grants EP/X012557/1 (D.T.S.P. and J.J.B.), EP/X01245X/1 (B.C. and E.K.)  and  the German Research Foundation (DFG) through CRC TRR 288 "Elasto-Q-Mat", project A07 (J.S.). The calculations were performed using King's Computational Research, Engineering and Technology Environment (CREATE).

\bibliography{References}

@article{VanHove1953,
  title = {{The Occurrence of Singularities in the Elastic Frequency Distribution of a Crystal}},
  author = {Van Hove, L\'eon},
  journal = {Phys. Rev.},
  volume = {89},
  issue = {6},
  pages = {1189--1193},
  numpages = {0},
  year = {1953},
  month = {Mar},
  publisher = {American Physical Society},
  doi = {10.1103/PhysRev.89.1189},
  url = {https://link.aps.org/doi/10.1103/PhysRev.89.1189}
}

@article{McChesney2010,
  title = {{Extended van Hove Singularity and Superconducting Instability in Doped Graphene}},
  author = {McChesney, J. L. and Bostwick, Aaron and Ohta, Taisuke and Seyller, Thomas and Horn, Karsten and Gonz\'alez, J. and Rotenberg, Eli},
  journal = {Phys. Rev. Lett.},
  volume = {104},
  issue = {13},
  pages = {136803},
  numpages = {4},
  year = {2010},
  month = {Apr},
  publisher = {American Physical Society},
  doi = {10.1103/PhysRevLett.104.136803},
  url = {https://link.aps.org/doi/10.1103/PhysRevLett.104.136803}
}

@article{Nandkishore2012,
  title={{Chiral superconductivity from repulsive interactions in doped graphene}},
  author={Nandkishore, Rahul and Levitov, Leonid S and Chubukov, Andrey V},
  journal={Nature Physics},
  volume={8},
  number={2},
  pages={158--163},
  year={2012},
  publisher={Nature Publishing Group UK London},
   url={https://doi.org/10.1038/nphys2208}
}

@article{Kiesel2012,
  title = {{Competing many-body instabilities and unconventional superconductivity in graphene}},
  author = {Kiesel, Maximilian L. and Platt, Christian and Hanke, Werner and Abanin, Dmitry A. and Thomale, Ronny},
  journal = {Phys. Rev. B},
  volume = {86},
  issue = {2},
  pages = {020507},
  numpages = {5},
  year = {2012},
  month = {Jul},
  publisher = {American Physical Society},
  doi = {10.1103/PhysRevB.86.020507},
  url = {https://link.aps.org/doi/10.1103/PhysRevB.86.020507}
}

@article{Wang2012,
  title = {{Functional renormalization group and variational Monte Carlo studies of the electronic instabilities in graphene near $\frac{1}{4}$ doping}},
  author = {Wang, Wan-Sheng and Xiang, Yuan-Yuan and Wang, Qiang-Hua and Wang, Fa and Yang, Fan and Lee, Dung-Hai},
  journal = {Phys. Rev. B},
  volume = {85},
  issue = {3},
  pages = {035414},
  numpages = {6},
  year = {2012},
  month = {Jan},
  publisher = {American Physical Society},
  doi = {10.1103/PhysRevB.85.035414},
  url = {https://link.aps.org/doi/10.1103/PhysRevB.85.035414}
}

@article{BlackSchaffer2014,
  title={{Chiral d-wave superconductivity in doped graphene}},
  author={Black-Schaffer, Annica M and Honerkamp, Carsten},
  journal={Journal of Physics: Condensed Matter},
  volume={26},
  number={42},
  pages={423201},
  year={2014},
  publisher={IOP Publishing},
  url={https://doi.org/10.1088/0953-8984/26/42/423201}
}

@article{Shtyk2017,
  title = {{Electrons at the monkey saddle: A multicritical Lifshitz point}},
  author = {Shtyk, A. and Goldstein, G. and Chamon, C.},
  journal = {Phys. Rev. B},
  volume = {95},
  issue = {3},
  pages = {035137},
  numpages = {10},
  year = {2017},
  month = {Jan},
  publisher = {American Physical Society},
  doi = {10.1103/PhysRevB.95.035137},
  url = {https://link.aps.org/doi/10.1103/PhysRevB.95.035137}
}

@article{Yuan2019,
  title={{Magic of high-order van Hove singularity}},
  author={Yuan, Noah FQ and Isobe, Hiroki and Fu, Liang},
  journal={Nature Communications},
  volume={10},
  number={1},
  pages={5769},
  year={2019},
  publisher={Nature Publishing Group UK London},
  url={https://doi.org/10.1038/s41467-019-13670-9}
}

@article{Isobe2019,
  title = {Supermetal},
  author = {Isobe, Hiroki and Fu, Liang},
  journal = {Phys. Rev. Res.},
  volume = {1},
  issue = {3},
  pages = {033206},
  numpages = {30},
  year = {2019},
  month = {Dec},
  publisher = {American Physical Society},
  doi = {10.1103/PhysRevResearch.1.033206},
  url = {https://link.aps.org/doi/10.1103/PhysRevResearch.1.033206}
}

@article{Yuan2020,
  title = {{Classification of critical points in energy bands based on topology, scaling, and symmetry}},
  author = {Yuan, Noah F. Q. and Fu, Liang},
  journal = {Phys. Rev. B},
  volume = {101},
  issue = {12},
  pages = {125120},
  numpages = {16},
  year = {2020},
  month = {Mar},
  publisher = {American Physical Society},
  doi = {10.1103/PhysRevB.101.125120},
  url = {https://link.aps.org/doi/10.1103/PhysRevB.101.125120}
}

@article{Hu2022,
  title={{Rich nature of Van Hove singularities in Kagome superconductor CsV${}_{3}$Sb${}_{5}$}},
  author={Hu, Yong and Wu, Xianxin and Ortiz, Brenden R and Ju, Sailong and Han, Xinloong and Ma, Junzhang and Plumb, Nicholas C and Radovic, Milan and Thomale, Ronny and Wilson, Stephen D and others},
  journal={Nature Communications},
  volume={13},
  number={1},
  pages={2220},
  year={2022},
  publisher={Nature Publishing Group UK London},
  url={https://doi.org/10.1038/s41467-022-29828-x}
}

@article{Efremov2019,
  title = {{Multicritical Fermi Surface Topological Transitions}},
  author = {Efremov, Dmitry V. and Shtyk, Alex and Rost, Andreas W. and Chamon, Claudio and Mackenzie, Andrew P. and Betouras, Joseph J.},
  journal = {Phys. Rev. Lett.},
  volume = {123},
  issue = {20},
  pages = {207202},
  numpages = {6},
  year = {2019},
  month = {Nov},
  publisher = {American Physical Society},
  doi = {10.1103/PhysRevLett.123.207202},
  url = {https://link.aps.org/doi/10.1103/PhysRevLett.123.207202}
}

@article{Chandrasekaran2020,
  title = {{Catastrophe theory classification of Fermi surface topological transitions in two dimensions}},
  author = {Chandrasekaran, A. and Shtyk, A. and Betouras, J. J. and Chamon, C.},
  journal = {Phys. Rev. Res.},
  volume = {2},
  issue = {1},
  pages = {013355},
  numpages = {19},
  year = {2020},
  month = {Mar},
  publisher = {American Physical Society},
  doi = {10.1103/PhysRevResearch.2.013355},
  url = {https://link.aps.org/doi/10.1103/PhysRevResearch.2.013355}
}

@article{Mousatov2020,
  title = {{Theory of the strange metal {Sr${}_{3}$Ru${}_{2}$O${}_{7}$}}},
  author = {Mousatov, C. H. and Berg, E. and Hartnoll, S. A.},
  journal = {Proc. Natl. Acad. Sci. U.S.A},
  volume = {117},
  issue = {6},
  pages = {2852},
  numpages = {6},
  year = {2020},
  month = {Jan},
  doi = {10.1073/pnas.1915224117},
  url = {https://doi.org/10.1073/pnas.1915224117}
}

@Article{Li2022,
author={Li, You-Sheng
and Garst, Markus
and Schmalian, J{\"o}rg
and Ghosh, Sayak
and Kikugawa, Naoki
and Sokolov, Dmitry A.
and Hicks, Clifford W.
and Jerzembeck, Fabian
and Ikeda, Matthias S.
and Hu, Zhenhai
and Ramshaw, B. J.
and Rost, Andreas W.
and Nicklas, Michael
and Mackenzie, Andrew P.},
title={{Elastocaloric determination of the phase diagram of Sr${}_{2}$RuO${}_{4}$}},
journal={Nature},
year={2022},
month={Jul},
day={01},
volume={607},
number={7918},
pages={276-280},
issn={1476-4687},
doi={10.1038/s41586-022-04820-z},
url={https://doi.org/10.1038/s41586-022-04820-z}
}

@article{Mackenzie2003,
  title = {{The superconductivity of ${\mathrm{Sr}}_{2}{\mathrm{RuO}}_{4}$ and the physics of spin-triplet pairing}},
  author = {Mackenzie, Andrew Peter and Maeno, Yoshiteru},
  journal = {Rev. Mod. Phys.},
  volume = {75},
  issue = {2},
  pages = {657--712},
  numpages = {0},
  year = {2003},
  month = {May},
  publisher = {American Physical Society},
  doi = {10.1103/RevModPhys.75.657},
  url = {https://link.aps.org/doi/10.1103/RevModPhys.75.657}
}

@article{Hussey1998,
  title = {{Normal-state magnetoresistance of ${\mathrm{Sr}}_{2}{\mathrm{RuO}}_{4}$}},
  author = {Hussey, N. E. and Mackenzie, A. P. and Cooper, J. R. and Maeno, Y. and Nishizaki, S. and Fujita, T.},
  journal = {Phys. Rev. B},
  volume = {57},
  issue = {9},
  pages = {5505--5511},
  numpages = {0},
  year = {1998},
  month = {Mar},
  publisher = {American Physical Society},
  doi = {10.1103/PhysRevB.57.5505},
  url = {https://link.aps.org/doi/10.1103/PhysRevB.57.5505}
}

@article{Katsufuji1996,
  title = {{In-Plane and Out-of-Plane Optical Spectra of ${\mathrm{Sr}}_{2}{\mathrm{RuO}}_{4}$}},
  author = {Katsufuji, T. and Kasai, M. and Tokura, Y.},
  journal = {Phys. Rev. Lett.},
  volume = {76},
  issue = {1},
  pages = {126--129},
  numpages = {0},
  year = {1996},
  month = {Jan},
  publisher = {American Physical Society},
  doi = {10.1103/PhysRevLett.76.126},
  url = {https://link.aps.org/doi/10.1103/PhysRevLett.76.126}
}

@article{Barber2019,
  title = {{Role of correlations in determining the Van Hove strain in ${\mathrm{Sr}}_{2}{\mathrm{RuO}}_{4}$}},
  author = {Barber, Mark E. and Lechermann, Frank and Streltsov, Sergey V. and Skornyakov, Sergey L. and Ghosh, Sayak and Ramshaw, B. J. and Kikugawa, Naoki and Sokolov, Dmitry A. and Mackenzie, Andrew P. and Hicks, Clifford W. and Mazin, I. I.},
  journal = {Phys. Rev. B},
  volume = {100},
  issue = {24},
  pages = {245139},
  numpages = {7},
  year = {2019},
  month = {Dec},
  publisher = {American Physical Society},
  doi = {10.1103/PhysRevB.100.245139},
  url = {https://link.aps.org/doi/10.1103/PhysRevB.100.245139}
}

@article{Hodges1971,
  title = {{Effect of Fermi Surface Geometry on Electron-Electron Scattering}},
  author = {Hodges, Christopher and Smith, Henrik and Wilkins, J. W.},
  journal = {Phys. Rev. B},
  volume = {4},
  issue = {2},
  pages = {302--311},
  numpages = {0},
  year = {1971},
  month = {Jul},
  publisher = {American Physical Society},
  doi = {10.1103/PhysRevB.4.302},
  url = {https://link.aps.org/doi/10.1103/PhysRevB.4.302}
}

@article{Bloom1975,
  title = {{Two-dimensional Fermi gas}},
  author = {Bloom, Paul},
  journal = {Phys. Rev. B},
  volume = {12},
  issue = {1},
  pages = {125--129},
  numpages = {0},
  year = {1975},
  month = {Jul},
  publisher = {American Physical Society},
  doi = {10.1103/PhysRevB.12.125},
  url = {https://link.aps.org/doi/10.1103/PhysRevB.12.125}
}

@article{Ojajarvi2024,
  title={{Pairing at a single Van Hove point}},
  author={Ojaj{\"a}rvi, Risto and Chubukov, Andrey V and Lee, Yueh-Chen and Garst, Markus and Schmalian, J{\"o}rg},
  journal={npj Quantum Materials},
  volume={9},
  number={1},
  pages={105},
  year={2024},
  publisher={Nature Publishing Group UK London},
   url={https://doi.org/10.1038/s41535-024-00717-4}
}

@article{Sidis1999,
  title = {{Evidence for Incommensurate Spin Fluctuations in ${\mathrm{Sr}}_{2}{\mathrm{RuO}}_{4}$}},
  author = {Sidis, Y. and Braden, M. and Bourges, P. and Hennion, B. and NishiZaki, S. and Maeno, Y. and Mori, Y.},
  journal = {Phys. Rev. Lett.},
  volume = {83},
  issue = {16},
  pages = {3320--3323},
  numpages = {0},
  year = {1999},
  month = {Oct},
  publisher = {American Physical Society},
  doi = {10.1103/PhysRevLett.83.3320},
  url = {https://link.aps.org/doi/10.1103/PhysRevLett.83.3320}
}

@article{Braden2002,
  title = {{Inelastic neutron scattering study of magnetic excitations in ${\mathrm{Sr}}_{2}{\mathrm{RuO}}_{4}$}},
  author = {Braden, M. and Sidis, Y. and Bourges, P. and Pfeuty, P. and Kulda, J. and Mao, Z. and Maeno, Y.},
  journal = {Phys. Rev. B},
  volume = {66},
  issue = {6},
  pages = {064522},
  numpages = {11},
  year = {2002},
  month = {Aug},
  publisher = {American Physical Society},
  doi = {10.1103/PhysRevB.66.064522},
  url = {https://link.aps.org/doi/10.1103/PhysRevB.66.064522}
}

@article{Palle2023,
  title = {{Constraints on the superconducting state of ${\text{Sr}}_{2}{\text{RuO}}_{4}$ from elastocaloric measurements}},
  author = {Palle, Grgur and Hicks, Clifford and Valent\'{\i}, Roser and Hu, Zhenhai and Li, You-Sheng and Rost, Andreas and Nicklas, Michael and Mackenzie, Andrew P. and Schmalian, J\"org},
  journal = {Phys. Rev. B},
  volume = {108},
  issue = {9},
  pages = {094516},
  numpages = {20},
  year = {2023},
  month = {Sep},
  publisher = {American Physical Society},
  doi = {10.1103/PhysRevB.108.094516},
  url = {https://link.aps.org/doi/10.1103/PhysRevB.108.094516}
}

@article{Bergemann2003,
  title={{Quasi-two-dimensional Fermi liquid properties of the unconventional superconductor Sr$_2$RuO$_4$}},
  author={Bergemann, C and Mackenzie, Andrew Peter and Julian, SR and Forsythe, D and Ohmichi, E},
  journal={Advances in Physics},
  volume={52},
  number={7},
  pages={639--725},
  year={2003},
  publisher={Taylor \& Francis},
  url={https://doi.org/10.1080/00018730310001621737}
}

@article{Lifshitz1960,
  title={{Anomalies of electron characteristics of a metal in the high pressure region}},
  author={Lifshitz, I. M.},
  journal={Sov. Phys. JETP},
  volume={11},
  number={5},
  pages={1130--1135},
  year={1960},
  url={http://jetp.ras.ru/cgi-bin/dn/e_011_05_1130.pdf}
}

@article{Kiesel2012B,
  title = {{Sublattice Interference in the Kagome {{Hubbard}} Model}},
  author = {Kiesel, Maximilian L. and Thomale, Ronny},
  year = {2012},
  month = sep,
  journal = {Physical Review B},
  volume = {86},
  number = {12},
  pages = {121105},
  issn = {1098-0121, 1550-235X},
  doi = {10.1103/PhysRevB.86.121105},
  urldate = {2023-12-01},
  langid = {english}
}

@article{Park2021,
  title = {{Electronic Instabilities of Kagome Metals: {{Saddle}} Points and {{Landau}} Theory}},
  shorttitle = {Electronic Instabilities of Kagome Metals},
  author = {Park, Takamori and Ye, Mengxing and Balents, Leon},
  year = {2021},
  month = jul,
  journal = {Physical Review B},
  volume = {104},
  number = {3},
  pages = {035142},
  issn = {2469-9950, 2469-9969},
  doi = {10.1103/PhysRevB.104.035142},
  urldate = {2024-08-19},
  langid = {english}
}

@article{Li2024,
  title = {{Intertwined {{Van Hove Singularities}} as a {{Mechanism}} for {{Loop Current Order}} in {{Kagome Metals}}}},
  author = {Li, Heqiu and Kim, Yong Baek and Kee, Hae-Young},
  year = {2024},
  month = apr,
  journal = {Physical Review Letters},
  volume = {132},
  number = {14},
  pages = {146501},
  issn = {0031-9007, 1079-7114},
  doi = {10.1103/PhysRevLett.132.146501},
  urldate = {2024-05-13},
  langid = {english}
}

@article{Servant2002,
  title = {{Magnetic excitations in the normal and superconducting states of ${\mathrm{Sr}}_{2}{\mathrm{RuO}}_{4}$}},
  author = {Servant, F. and F\aa{}k, B. and Raymond, S. and Brison, J. P. and Lejay, P. and Flouquet, J.},
  journal = {Phys. Rev. B},
  volume = {65},
  issue = {18},
  pages = {184511},
  numpages = {6},
  year = {2002},
  month = {Apr},
  publisher = {American Physical Society},
  doi = {10.1103/PhysRevB.65.184511},
  url = {https://link.aps.org/doi/10.1103/PhysRevB.65.184511}
}

@article{Braden2004,
  title = {{Anisotropy of the Incommensurate Fluctuations in ${\mathrm{S}\mathrm{r}}_{2}{\mathrm{R}\mathrm{u}\mathrm{O}}_{4}$: A Study with Polarized Neutrons}},
  author = {Braden, M. and Steffens, P. and Sidis, Y. and Kulda, J. and Bourges, P. and Hayden, S. and Kikugawa, N. and Maeno, Y.},
  journal = {Phys. Rev. Lett.},
  volume = {92},
  issue = {9},
  pages = {097402},
  numpages = {4},
  year = {2004},
  month = {Mar},
  publisher = {American Physical Society},
  doi = {10.1103/PhysRevLett.92.097402},
  url = {https://link.aps.org/doi/10.1103/PhysRevLett.92.097402}
}

@article{Iida2011,
  title = {{Inelastic neutron scattering study of the magnetic fluctuations in Sr${}_{2}$RuO${}_{4}$}},
  author = {Iida, K. and Kofu, M. and Katayama, N. and Lee, J. and Kajimoto, R. and Inamura, Y. and Nakamura, M. and Arai, M. and Yoshida, Y. and Fujita, M. and Yamada, K. and Lee, S.-H.},
  journal = {Phys. Rev. B},
  volume = {84},
  issue = {6},
  pages = {060402},
  numpages = {4},
  year = {2011},
  month = {Aug},
  publisher = {American Physical Society},
  doi = {10.1103/PhysRevB.84.060402},
  url = {https://link.aps.org/doi/10.1103/PhysRevB.84.060402}
}

@article{Jenni2021,
  title = {{Neutron scattering studies on spin fluctuations in ${\mathrm{Sr}}_{2}{\mathrm{RuO}}_{4}$}},
  author = {Jenni, K. and Kunkem\"oller, S. and Steffens, P. and Sidis, Y. and Bewley, R. and Mao, Z. Q. and Maeno, Y. and Braden, M.},
  journal = {Phys. Rev. B},
  volume = {103},
  issue = {10},
  pages = {104511},
  numpages = {10},
  year = {2021},
  month = {Mar},
  publisher = {American Physical Society},
  doi = {10.1103/PhysRevB.103.104511},
  url = {https://link.aps.org/doi/10.1103/PhysRevB.103.104511}
}

@article{Burganov2016,
  title = {{Strain Control of Fermiology and Many-Body Interactions in Two-Dimensional Ruthenates}},
  author = {Burganov, B. and Adamo, C. and Mulder, A. and Uchida, M. and King, P. D. C. and Harter, J. W. and Shai, D. E. and Gibbs, A. S. and Mackenzie, A. P. and Uecker, R. and Bruetzam, M. and Beasley, M. R. and Fennie, C. J. and Schlom, D. G. and Shen, K. M.},
  journal = {Phys. Rev. Lett.},
  volume = {116},
  issue = {19},
  pages = {197003},
  numpages = {6},
  year = {2016},
  month = {May},
  publisher = {American Physical Society},
  doi = {10.1103/PhysRevLett.116.197003},
  url = {https://link.aps.org/doi/10.1103/PhysRevLett.116.197003}
}

@article{Coffey1993,
  title = {{Nonanalytic contributions to the self-energy and the thermodynamics of two-dimensional Fermi liquids}},
  author = {Coffey, D. and Bedell, K. S.},
  journal = {Phys. Rev. Lett.},
  volume = {71},
  issue = {7},
  pages = {1043--1046},
  numpages = {0},
  year = {1993},
  month = {Aug},
  publisher = {American Physical Society},
  doi = {10.1103/PhysRevLett.71.1043},
  url = {https://link.aps.org/doi/10.1103/PhysRevLett.71.1043}
}

@article{Kugler2020,
  title = {{Strongly Correlated Materials from a Numerical Renormalization Group Perspective: How the Fermi-Liquid State of ${\mathrm{Sr}}_{2}{\mathrm{RuO}}_{4}$ Emerges}},
  author = {Kugler, Fabian B. and Zingl, Manuel and Strand, Hugo U. R. and Lee, Seung-Sup B. and von Delft, Jan and Georges, Antoine},
  journal = {Phys. Rev. Lett.},
  volume = {124},
  issue = {1},
  pages = {016401},
  numpages = {7},
  year = {2020},
  month = {Jan},
  publisher = {American Physical Society},
  doi = {10.1103/PhysRevLett.124.016401},
  url = {https://link.aps.org/doi/10.1103/PhysRevLett.124.016401}
}

@article{Barber2018,
  title = {{Resistivity in the Vicinity of a Van Hove Singularity: ${\mathrm{Sr}}_{2}{\mathrm{RuO}}_{4}$ under Uniaxial Pressure}},
  author = {Barber, M. E. and Gibbs, A. S. and Maeno, Y. and Mackenzie, A. P. and Hicks, C. W.},
  journal = {Phys. Rev. Lett.},
  volume = {120},
  issue = {7},
  pages = {076602},
  numpages = {6},
  year = {2018},
  month = {Feb},
  publisher = {American Physical Society},
  doi = {10.1103/PhysRevLett.120.076602},
  url = {https://link.aps.org/doi/10.1103/PhysRevLett.120.076602}
}

@article{Noad2023,
  title={{Giant lattice softening at a Lifshitz transition in Sr${}_{2}$RuO${}_{4}$}},
  author={Noad, Hilary ML and Ishida, Kousuke and Li, Y-S and Gati, Elena and Stangier, Veronika and Kikugawa, Naoki and Sokolov, Dmitry A and Nicklas, Michael and Kim, Bongjae and Mazin, Igor I and others},
  journal={Science},
  volume={382},
  number={6669},
  pages={447--450},
  year={2023},
  publisher={American Association for the Advancement of Science},
   url={https://doi.org/10.1126/science.adf3348}
}

@article{Zingl2019,
  title={{Hall coefficient signals orbital differentiation in the Hund’s metal Sr${}_{2}$RuO${}_{4}$}},
  author={Zingl, Manuel and Mravlje, Jernej and Aichhorn, Markus and Parcollet, Olivier and Georges, Antoine},
  journal={npj Quantum Materials},
  volume={4},
  number={1},
  pages={35},
  year={2019},
  publisher={Nature Publishing Group UK London},
  url={https://doi.org/10.1038/s41535-019-0175-y}
}

@article{Suzuki2023,
  title={{Distinct spin and orbital dynamics in Sr${}_{2}$RuO${}_{4}$}},
  author={Suzuki, Hakuto and Wang, L and Bertinshaw, J and Strand, Hugo UR and K{\"a}ser, S and Krautloher, Maximilian and Yang, Z and Wentzell, N and Parcollet, O and Jerzembeck, F and others},
  journal={Nature Communications},
  volume={14},
  number={1},
  pages={7042},
  year={2023},
  publisher={Nature Publishing Group UK London},
   url={https://doi.org/10.1038/s41467-023-42804-3
}
}

@article{Gopalan1992,
  title = {{Effects of saddle-point singularities on the electron lifetime}},
  author = {Gopalan, S. and Gunnarsson, O. and Andersen, O. K.},
  journal = {Phys. Rev. B},
  volume = {46},
  issue = {18},
  pages = {11798--11806},
  numpages = {0},
  year = {1992},
  month = {Nov},
  publisher = {American Physical Society},
  doi = {10.1103/PhysRevB.46.11798},
  url = {https://link.aps.org/doi/10.1103/PhysRevB.46.11798}
}

@article{Pattnaik1992,
  title = {{Evidence for the van Hove scenario in high-temperature superconductivity from quasiparticle-lifetime broadening}},
  author = {Pattnaik, P. C. and Kane, C. L. and Newns, D. M. and Tsuei, C. C.},
  journal = {Phys. Rev. B},
  volume = {45},
  issue = {10},
  pages = {5714--5717},
  numpages = {0},
  year = {1992},
  month = {Mar},
  publisher = {American Physical Society},
  doi = {10.1103/PhysRevB.45.5714},
  url = {https://link.aps.org/doi/10.1103/PhysRevB.45.5714}
}

@article{Yang2023,
  title = {{Probing Momentum-Dependent Scattering in Uniaxially Stressed ${\mathrm{Sr}}_{2}{\mathrm{RuO}}_{4}$ through the Hall Effect}},
  author = {Yang, Po-Ya and Noad, Hilary M. L. and Barber, Mark E. and Kikugawa, Naoki and Sokolov, Dmitry A. and Mackenzie, Andrew P. and Hicks, Clifford W.},
  journal = {Phys. Rev. Lett.},
  volume = {131},
  issue = {3},
  pages = {036301},
  numpages = {6},
  year = {2023},
  month = {Jul},
  publisher = {American Physical Society},
  doi = {10.1103/PhysRevLett.131.036301},
  url = {https://link.aps.org/doi/10.1103/PhysRevLett.131.036301}
}

@Article{Steppke2017,
author={Steppke, Alexander
and Zhao, Lishan
and Barber, Mark E.
and Scaffidi, Thomas
and Jerzembeck, Fabian
and Rosner, Helge
and Gibbs, Alexandra S.
and Maeno, Yoshiteru
and Simon, Steven H.
and Mackenzie, Andrew P.
and Hicks, Clifford W.},
title={{Strong peak in $T_c$ of Sr$_2$RuO$_4$ under uniaxial pressure}},
journal={Science},
year={2017},
month={Jan},
day={13},
publisher={American Association for the Advancement of Science},
volume={355},
number={6321},
pages={eaaf9398},
doi={10.1126/science.aaf9398},
url={https://doi.org/10.1126/science.aaf9398}
}

@article{Hicks2014,
author = {Clifford W. Hicks  and Daniel O. Brodsky  and Edward A. Yelland  and Alexandra S. Gibbs  and Jan A. N. Bruin  and Mark E. Barber  and Stephen D. Edkins  and Keigo Nishimura  and Shingo Yonezawa  and Yoshiteru Maeno  and Andrew P. Mackenzie },
title = {{Strong Increase of $T_c$ of Sr$_2$RuO$_4$ Under Both Tensile and Compressive Strain}},
journal = {Science},
volume = {344},
number = {6181},
pages = {283-285},
year = {2014},
doi = {10.1126/science.1248292},
URL = {https://www.science.org/doi/abs/10.1126/science.1248292},
eprint = {https://www.science.org/doi/pdf/10.1126/science.1248292}}

@misc{Hunter2025,
      title={{Non-Fermi liquid quasiparticles in strain-tuned Sr${}_{2}$RuO${}_{4}$}}, 
      author={A. Hunter and C. Putzke and F. B. Kugler and S. Beck and E. Cappelli and F. Margot and M. Straub and Y. Alexanian and J. Teyssier and A. de la Torre and K. W. Plumb and M. D. Watson and T. K. Kim and C. Cacho and N. C. Plumb and M. Shi and M. Radovic and J. Osiecki and C. Polley and D. A. Sokolov and A. P. Mackenzie and E. Berg and A. Georges and P. J. W. Moll and A. Tamai and F. Baumberger},
      year={2025},
      eprint={2503.11311},
      archivePrefix={arXiv},
      primaryClass={cond-mat.str-el},
      url={https://arxiv.org/abs/2503.11311}, 
}

@Article{Ghosh2021,
author={Ghosh, Sayak
and Shekhter, Arkady
and Jerzembeck, F.
and Kikugawa, N.
and Sokolov, Dmitry A.
and Brando, Manuel
and Mackenzie, A. P.
and Hicks, Clifford W.
and Ramshaw, B. J.},
title={{Thermodynamic evidence for a two-component superconducting order parameter in Sr${}_{2}$RuO${}_{4}$}},
journal={Nature Physics},
year={2021},
month={Feb},
day={01},
volume={17},
number={2},
pages={199-204},
issn={1745-2481},
doi={10.1038/s41567-020-1032-4},
url={https://doi.org/10.1038/s41567-020-1032-4}
}

@Article{Sunko2019,
author={Sunko, Veronika
and Abarca Morales, Edgar
and Markovi{\'{c}}, Igor
and Barber, Mark E.
and Milosavljevi{\'{c}}, Dijana
and Mazzola, Federico
and Sokolov, Dmitry A.
and Kikugawa, Naoki
and Cacho, Cephise
and Dudin, Pavel
and Rosner, Helge
and Hicks, Clifford W.
and King, Philip D. C.
and Mackenzie, Andrew P.},
title={{Direct observation of a uniaxial stress-driven Lifshitz transition in Sr${}_{2}$RuO${}_{4}$}},
journal={npj Quantum Materials},
year={2019},
month={Aug},
day={19},
volume={4},
number={1},
pages={46},
issn={2397-4648},
doi={10.1038/s41535-019-0185-9},
url={https://doi.org/10.1038/s41535-019-0185-9}
}

@article{Barber2019_correlations,
  title = {{Role of correlations in determining the Van Hove strain in ${\mathrm{Sr}}_{2}{\mathrm{RuO}}_{4}$}},
  author = {Barber, Mark E. and Lechermann, Frank and Streltsov, Sergey V. and Skornyakov, Sergey L. and Ghosh, Sayak and Ramshaw, B. J. and Kikugawa, Naoki and Sokolov, Dmitry A. and Mackenzie, Andrew P. and Hicks, Clifford W. and Mazin, I. I.},
  journal = {Phys. Rev. B},
  volume = {100},
  issue = {24},
  pages = {245139},
  numpages = {7},
  year = {2019},
  month = {Dec},
  publisher = {American Physical Society},
  doi = {10.1103/PhysRevB.100.245139},
  url = {https://link.aps.org/doi/10.1103/PhysRevB.100.245139}
}

@article{Hlubina1996,
  title = {Effect of impurities on the transport properties in the Van Hove scenario},
  author = {Hlubina, R.},
  journal = {Phys. Rev. B},
  volume = {53},
  issue = {17},
  pages = {11344--11347},
  numpages = {0},
  year = {1996},
  month = {May},
  publisher = {American Physical Society},
  doi = {10.1103/PhysRevB.53.11344},
  url = {https://link.aps.org/doi/10.1103/PhysRevB.53.11344}
}

@article{Mackenzie1996,
  title = {Quantum Oscillations in the Layered Perovskite Superconductor S${\mathrm{r}}_{2}$Ru${\mathrm{O}}_{4}$},
  author = {Mackenzie, A. P. and Julian, S. R. and Diver, A. J. and McMullan, G. J. and Ray, M. P. and Lonzarich, G. G. and Maeno, Y. and Nishizaki, S. and Fujita, T.},
  journal = {Phys. Rev. Lett.},
  volume = {76},
  issue = {20},
  pages = {3786--3789},
  numpages = {0},
  year = {1996},
  month = {May},
  publisher = {American Physical Society},
  doi = {10.1103/PhysRevLett.76.3786},
  url = {https://link.aps.org/doi/10.1103/PhysRevLett.76.3786}
}

@article{Dzyaloshinskii1996,
  title={Extended Van-Hove singularity and related non-Fermi liquids},
  author={Dzyaloshinskii, Igor},
  journal={Journal de Physique I},
  volume={6},
  number={1},
  pages={119--135},
  year={1996},
  publisher={EDP Sciences},
  url = {https://doi.org/10.1051/jp1:1996127}
}

@article{Menashe1999,
  title = {Fermi-liquid properties of a two-dimensional electron system with the Fermi level near a van Hove singularity},
  author = {Menashe, D. and Laikhtman, B.},
  journal = {Phys. Rev. B},
  volume = {59},
  issue = {21},
  pages = {13592--13595},
  numpages = {0},
  year = {1999},
  month = {Jun},
  publisher = {American Physical Society},
  doi = {10.1103/PhysRevB.59.13592},
  url = {https://link.aps.org/doi/10.1103/PhysRevB.59.13592}
}

@article{Ingle2005,
  title = {Quantitative analysis of ${\mathrm{Sr}}_{2}{\mathrm{RuO}}_{4}$ angle-resolved photoemission spectra: Many-body interactions in a model Fermi liquid},
  author = {Ingle, N. J. C. and Shen, K. M. and Baumberger, F. and Meevasana, W. and Lu, D. H. and Shen, Z.-X. and Damascelli, A. and Nakatsuji, S. and Mao, Z. Q. and Maeno, Y. and Kimura, T. and Tokura, Y.},
  journal = {Phys. Rev. B},
  volume = {72},
  issue = {20},
  pages = {205114},
  numpages = {9},
  year = {2005},
  month = {Nov},
  publisher = {American Physical Society},
  doi = {10.1103/PhysRevB.72.205114},
  url = {https://link.aps.org/doi/10.1103/PhysRevB.72.205114}
}

@article{Iwasawa2010,
  title = {{Interplay among Coulomb Interaction, Spin-Orbit Interaction, and Multiple Electron-Boson Interactions in Sr${}_{2}$RuO${}_{4}$}},
  author = {Iwasawa, H. and Yoshida, Y. and Hase, I. and Koikegami, S. and Hayashi, H. and Jiang, J. and Shimada, K. and Namatame, H. and Taniguchi, M. and Aiura, Y.},
  journal = {Phys. Rev. Lett.},
  volume = {105},
  issue = {22},
  pages = {226406},
  numpages = {4},
  year = {2010},
  month = {Nov},
  publisher = {American Physical Society},
  doi = {10.1103/PhysRevLett.105.226406},
  url = {https://link.aps.org/doi/10.1103/PhysRevLett.105.226406}
}

@article{Tamai2019,
  title = {High-Resolution Photoemission on ${\mathrm{Sr}}_{2}{\mathrm{RuO}}_{4}$ Reveals Correlation-Enhanced Effective Spin-Orbit Coupling and Dominantly Local Self-Energies},
  author = {Tamai, A. and Zingl, M. and Rozbicki, E. and Cappelli, E. and Ricc\`o, S. and de la Torre, A. and McKeown Walker, S. and Bruno, F. Y. and King, P. D. C. and Meevasana, W. and Shi, M. and Radovi\ifmmode \acute{c}\else \'{c}\fi{}, M. and Plumb, N. C. and Gibbs, A. S. and Mackenzie, A. P. and Berthod, C. and Strand, H. U. R. and Kim, M. and Georges, A. and Baumberger, F.},
  journal = {Phys. Rev. X},
  volume = {9},
  issue = {2},
  pages = {021048},
  numpages = {18},
  year = {2019},
  month = {Jun},
  publisher = {American Physical Society},
  doi = {10.1103/PhysRevX.9.021048},
  url = {https://link.aps.org/doi/10.1103/PhysRevX.9.021048}
}

@article{Stangier2022,
  title = {{Breakdown of the Wiedemann-Franz law at the Lifshitz point of strained ${\mathrm{Sr}}_{2}{\mathrm{RuO}}_{4}$}},
  author = {Stangier, Veronika C. and Berg, Erez and Schmalian, J\"org},
  journal = {Phys. Rev. B},
  volume = {105},
  issue = {11},
  pages = {115113},
  numpages = {12},
  year = {2022},
  month = {Mar},
  publisher = {American Physical Society},
  doi = {10.1103/PhysRevB.105.115113},
  url = {https://link.aps.org/doi/10.1103/PhysRevB.105.115113}
}

@article{Chandrasekaran2024,
  title = {{On the engineering of higher-order Van Hove singularities in two dimensions}},
  author = {Chandrasekaran, A. and Luke, R. C. and Morales, E. A. and Marques, C. A. and King, P. D. C. and Wahl, P. and Betouras, J. J.},
  journal = {Nature Communications},
  volume = {15},
  issue = {1},
  pages = {9521},
  year = {2024},
  month = {Nov},
  publisher = {Nature},
  doi = {https://doi.org/10.1038/s41467-024-53650-2},
  url = {https://doi.org/10.1038/s41467-021-27042-9},
}

@article{Classen_Betouras_2025,
  title={{High-order Van Hove singularities and their connection to flat bands}},
  author={Classen, Laura and Betouras, Joseph J},
  journal={Annual Review of Condensed Matter},
  volume={16},
  pages={229},
  year={2025},
   url={https://doi.org/10.1146/annurev-conmatphys-042924-015000}
}

@article{Damascelli2000,
  title = {{Fermi Surface, Surface States, and Surface Reconstruction in Sr${}_{2}$RuO${}_{4}$}},
  author = {Damascelli, A. and Lu, D. H. and Shen, K. M. and Armitage, N. P. and Ronning, F. and Feng, D. L. and Kim, C. and Shen, Z.-X. and Kimura, T. and Tokura, Y. and Mao, Z. Q. and Maeno, Y.},
  journal = {Phys. Rev. Lett.},
  volume = {85},
  issue = {24},
  pages = {5194--5197},
  numpages = {0},
  year = {2000},
  month = {Dec},
  publisher = {American Physical Society},
  doi = {10.1103/PhysRevLett.85.5194},
  url = {https://link.aps.org/doi/10.1103/PhysRevLett.85.5194}
}

\appendix
\section{On the non-monotonic hh$\to$hh scattering }
\label{app:AppendixA}

In the main text we  presented the curious result  that the scattering rate $\Gamma_{\boldsymbol{k}}\left(\omega,T\right)$
at the Lifshitz transition and for $\boldsymbol{k}$ at the Van Hove point is at finite
$T$ a non-monotonic function of frequency $\omega$. Here we confirm
this finding using an analytical approach. This analysis  shows that the
minimum of the scattering rate occurs for $\omega\sim T$ and is a result of competing tendency of thermal and quantum excitations of the compressive mode described by Eq.~\eqref{eq:Pi}.

For the analysis of hh $\to$ hh processes we can focus on the states in the vicinity of the Van Hove point and  use the dispersion
\begin{equation}
\varepsilon_{\boldsymbol{k}}=\frac{1}{2m}\left(k_{x}^{2}-k_{y}^{2}\right).
\end{equation}
The corresponding electronic
density of states is $\rho\left(\epsilon\right)=\frac{m}{\pi^{2}}\log\left(D/\left|\epsilon\right|\right)$.
$\sqrt{mD}$ is the typical momentum scale where the above expansion
is valid, i.e. approximately the region marked in red in Fig.~\ref{fig:hot_cold_and_scattering}. At
finite $T$ the spectrum $  {\rm Im}\Pi\left(\boldsymbol{q},\omega\right)$ of compressive modes can be described by  Eq.~\eqref{eq:Pi} with $\varepsilon_{{\rm VH}}\left(\boldsymbol{q}\right)$ replaced by $\sqrt{\varepsilon_{{\rm VH}}\left(\boldsymbol{q}\right)^2+T^2}$~\cite{Stangier2022}. We further note that with Eq.~\eqref{eq:Pi} the momentum dependence of $  {\rm Im}\Pi\left(\boldsymbol{q},\omega\right)$ only enters through $\varepsilon_{{\rm VH}}\left(\boldsymbol{q}\right)$ which we indicate by 
\begin{equation}
{\rm Im}\Pi\left(\boldsymbol{q},\omega \right)  =  {\rm Im}\Pi\left(\varepsilon_{{\rm VH}}\left(\boldsymbol{q}\right),\omega \right).
 \end{equation}
 The scattering rate right at the Van Hove point 
\begin{equation}
\Gamma_{{\rm VH}}\left(\omega,T\right)=\Gamma_{\boldsymbol{k}}\left(\omega,T\right)
\end{equation}
can then be written as an integral over energy only:
 \begin{equation}
\Gamma_{{\rm VH}}\left(\omega,T\right)  =  2U^{2}\int d\epsilon\rho\left(\epsilon\right) S\left(\epsilon,\omega \right) {\rm Im}\Pi\left(\epsilon,\omega-\epsilon\right),
\end{equation}
with  $S\left(\epsilon,\omega \right)=n_{F}\left(\epsilon\right)+n_{B}\left(\epsilon-\omega\right)$.

We first consider  the limit $T=0$. Then follows 
\begin{eqnarray}
\Gamma_{{\rm VH}}\left(\omega,0\right) & = & \frac{U^{2}m}{\pi}\left(\int_{0}^{\omega/2}d\epsilon\rho\left(\varepsilon\right)+\int_{\omega/2}^{\omega}d\epsilon\rho\left(\varepsilon\right)\frac{\omega-\epsilon}{\epsilon}\right)\nonumber \\
 & = & \lambda\omega\left(1+\log\frac{\sqrt{2}D}{\omega}\right),\label{eq:GVH_T0}
\end{eqnarray}
with dimensionless coupling constant $\lambda=m^{2}U^{2}\log2/\pi^{3}$. 
In the opposite limit of $\omega=0$ but finite $T$ it holds instead
\begin{eqnarray}
\Gamma_{{\rm VH}}\left(0,T\right) & = & \frac{2U^{2}m^{2}}{\pi^{3}}T\int_{0}^{D/T}dx\frac{\log\frac{D}{xT}}{\sinh\left(x\right)}\frac{x}{\sqrt{x^{2}+1}}\nonumber \\
 & = & \frac{2\lambda}{\log2}T\left(c+c'\log\frac{D}{T}\right),
 \label{eq:GVHomg0}
\end{eqnarray}
with $c=0.5338$ and $c'=1.49726$. The integrals over $x$
have been numerically evaluated with the upper limit $D/T\rightarrow\infty$. 

Finally, we consider the case with generic $\omega/T$:
\begin{eqnarray}
\Gamma_{{\rm VH}}\left(\omega,T\right) & = &-\frac{U^{2}m^{2}}{\pi^{3}}\int_{\epsilon^{*}}^{D}d\epsilon\log\left(\frac{D}{\left|\epsilon\right|}\right)\frac{S\left(\epsilon,\omega \right)\left(\omega-\epsilon\right)}{\sqrt{\epsilon^{2}+T^{2}}}  \nonumber \\
& - &\frac{U^{2}m^{2}}{\pi^{3}}\int_{-D}^{\epsilon^{*}}d\epsilon\log\left(\frac{D}{\left|\epsilon\right|}\right)S\left(\epsilon,\omega \right),
\end{eqnarray}
where $\epsilon^{*}=\frac{\omega^{2}-T^{2}}{2\omega}$. This can be
written as 
\begin{eqnarray}
\Gamma_{{\rm VH}}\left(\omega,T\right) & = & \lambda T\left(Q_{\omega}\left(\frac{\omega}{T}\right)+Q_{T}\left(\frac{\omega}{T}\right)\right)\label{eq:total_rate_min}
\end{eqnarray}
with 
\begin{eqnarray}
Q_{\omega}\left(y\right) & = & -\int_{-d}^{x^{*}\left(y\right)}\frac{dx\log\left(\frac{d}{\left|x\right|}\right)}{\log2} s\left(x,y\right),\nonumber \\
Q_{T}\left(y\right) & = & -\int_{x^{*}\left(y\right)}^{d}\frac{dx\log\left(\frac{d}{\left|x\right|}\right)}{\log2}\frac{s\left(x,y\right)\left(y-x\right)}{\sqrt{x^{2}+1}}.
\end{eqnarray}
Here, $x^{*}\left(y\right)=\frac{y^{2}-1}{2y}$ and
$s\left(x,y\right)=\frac{1}{e^{x}+1}+\frac{1}{e^{x-y}-1}$.
Notice, $Q_{\omega}$
and $Q_{T}$ also depend on the ratio of the upper energy cut off and temperature: $d\equiv D/T$. Except for the argument of the
logarithm, this dependency is weak and hence suppressed. 
The zero temperature
rate of Eq.(\ref{eq:GVH_T0}) can be written as $Q_{0}\left(y\right)=y\left(1+\log\frac{\sqrt{2}d}{y}\right)$,
while $Q\left(0\right)=\frac{2}{\log2}\left(c+c'\log d\right)$
corresponds to Eq.(\ref{eq:GVHomg0}). 

\begin{figure}[t]
    \centering
    \includegraphics[width=1\linewidth]{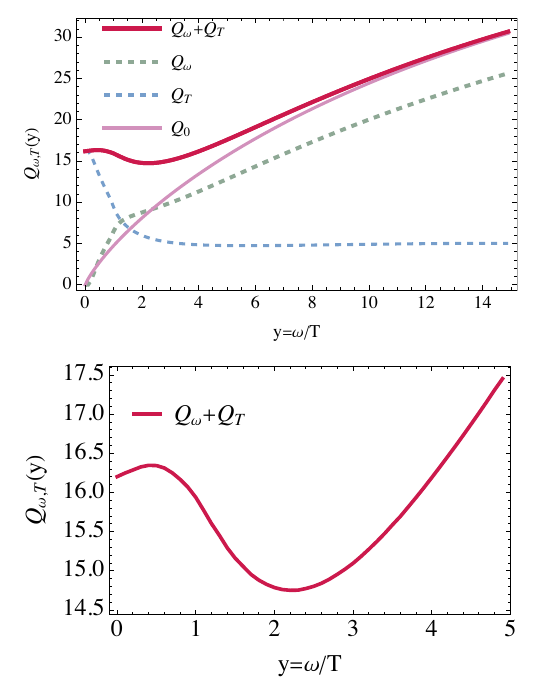}
    \caption{\justifying Quantum $Q_{\omega}$ and thermal $Q_{T}$ contributions to the scattering rate Eq.(\ref{eq:total_rate_min}) as function of $\omega/T$ along with their sum and in comparison with the zero-temperature contribution that follows from Eq.(\ref{eq:GVH_T0}). The lower panel shows $Q_{\omega}+Q_{T}$ near the minimum at $\omega\approx2.25T$. The minimum is a consequence of the rapid drop in $Q_{T}$ for energies comparable to $T$.}
    \label{fig:minimum_analytic}
\end{figure}

$Q_{\omega}$ is the \emph{quantum contribution} to the scattering rate and is caused by compressive excitations with energies larger than $T$, conversely $Q_{T}$ is, except for an overall constant, dominated by the corresponding \emph{thermal contribution}.  In Fig.~\ref{fig:minimum_analytic} we show the $y=\omega/T$-dependence of $Q_{\omega}\left(y\right)$, $Q_{T}\left(y\right)$ (dashed lines) and their sum (solid red line) in comparison with the $T=0$ result $Q_{0}\left(y\right)$ (pink solid line) at large $d=30$. $Q_{\omega}$ dominates at $\omega\gg T$ and gives rise to the  $\omega$$\log\omega$ dependence of the scattering rate.  Only $Q_{T}$ contributes at $\omega=0$. It approaches a constant at large $y$. For small but finite argument, the $y$-dependence of the two contributions is opposite and almost cancels to a constant. However, for $\omega\sim T$ the thermal contribution to the rate, governed by $Q_{T}$, drops  faster than the quantum contribution $Q_{\omega}$ rises. This is the reason for the local minimum of the scattering rate at $\omega\approx2.25T$. 

Finally, we comment on the detailed role of logarithmic corrections to the scattering rate $\Gamma$ and the inverse life time $\tau^{-1}$. At the VHS the dominant low-energy contribution to imaginary part of the self energy is ${\rm Im}\Sigma\left(\omega\right)=-\lambda\left\vert \omega\right\vert \log\frac{D}{\left\vert \omega\right\vert }$; see Eq.~\eqref{eq:GVH_T0}. Kramers-Kronig transforming this result  yields for the real part ${\rm Re}\Sigma\left(\omega\right)=-\frac{\lambda\omega}{\pi}\log^{2}\frac{D}{\left\vert \omega\right\vert}$. Hence the quasi-particle residue of Eq.~\eqref{eq:qpresidue} for small $\omega$ behaves as
$Z\left(\omega\right)=\frac{\pi}{\lambda\log^{2}\frac{D}{\omega}}$. Hence, we obtain for the inverse life time of Eq.~\eqref{eq:tau}: $ \tau^{-1}(\omega)=2\pi\left\vert \omega\right\vert /\log\frac{\Lambda}{\omega}$. We note that  the  logarithmic contribution to the self energy  is a direct consequence of the logarithmic divergence of the density of states at the VH point. If a constant background were included in the density of states, the relative magnitude of the coefficient of the logarithmic terms would be reduced, which explains why  they are harder to observe in the numerical analysis of the tight-binding dispersion presented in the main text. 

\end{document}